\documentclass[prb,twocolumn,showpacs,showkeys,superscriptaddress]{revtex4}

\usepackage{hyperref}
\usepackage[english]{babel}
\usepackage[T1]{fontenc}
\usepackage[latin1]{inputenc}
\usepackage{amsmath}
\usepackage{mathenv}
\usepackage{bm}
\usepackage{babel}
\usepackage[v2]{xy}
\usepackage{epsfig}
\usepackage{amsfonts}
\usepackage{amssymb}
\usepackage{color}
\usepackage{epstopdf} 

\begin{document}

\title{Tunability versus deviation sensitivity in a non-linear vortex oscillator}
\date{\today}
\author{S.Y. Martin}
\affiliation{SPINTEC, UMR-8191,CEA-INAC/CNRS/UJF-Grenoble 1/Grenoble-INP, 17 rue des martyrs, 38054 Grenoble Cedex 9, France}
\author{C. Thirion}
\affiliation{Institut Néel, CNRS et Université Joseph Fourier, BP 166, F-38042 Grenoble Cedex 9, France}
\author{C. Hoarau}
\affiliation{Institut Néel, CNRS et Université Joseph Fourier, BP 166, F-38042 Grenoble Cedex 9, France}
\author{C. Baraduc}
\author{B. Diény}
\affiliation{SPINTEC, UMR-8191,CEA-INAC/CNRS/UJF-Grenoble 1/Grenoble-INP, 17 rue des martyrs, 38054 Grenoble Cedex 9, France}

\begin{abstract}
Frequency modulation experiments were performed on a spin torque vortex oscillator for a wide range of modulation frequency, up to 10 \% of the oscillator frequency. A thorough analysis of the intermodulation products shows that the key parameter that describes these experiments is the deviation sensitivity, which is the dynamical frequency-current dependence. It  differs significantly from the oscillator tunability discussed so far in the context of spin-transfer oscillators. The essential difference between these two concepts is related to the response time of the vortex oscillator, driven either in quasi-steady state or in a transient regime.
\end{abstract}
\maketitle
\clearpage

\section{Introduction}
Spin-transfer oscillators are based on the excitation of magnetization precession by a large dc current density. These oscillators combine many interesting features, such as broad-band frequency operation, small size, easy integration and scalability. In this context, vortex oscillators are promising devices,  with  intrinsically  high output power and narrow linewidth \cite{Pufall_2007}.  The oscillation is due to the gyrotropic motion of a magnetic vortex set into motion by the spin-transfer torque. Such systems were initially studied with nanocontacts on spin valve structures \cite{Pufall_2007,Mistral_2008},
 and, later, in magnetic tunnel junction nanopillars \cite{Dussaux_2010}  under  an out-of-plane applied field.

When characterizing an oscillator for telecommunication applications, studying the modulation of the output (\textit{carrier}) by the information channel (\textit{modulation wave}) is essential to evaluate its potential. Moreover, such frequency modulation experiments lead to a better insight into the magnetization dynamics. Up to now, such studies were devoted only  to macrospin oscillators. Here we performed low-noise frequency modulation measurements on a vortex oscillator, based on a magnetic tunnel junction nanopillar in an in-plane field \cite{Martin_2011}. We also develop an analysis  based on modulation theory to treat the whole  set of data at once. It appears that the description previously used to describe modulation experiments where the   modulation frequency was much less than  the natural frequency, does not apply when the modulation frequency is  a significant fraction of the natural frequency. More precisely, the vortex response time appears to play a significant role, so that the concept of deviation sensitivity\cite{Golio_2010, Bock_1978} has to be introduced to explain the  observations. Until now, deviation sensitivity was overlooked in the case of spintronic oscillators: it corresponds to the dynamical dependence of the oscillator frequency with an applied  current that varies with time. We show that this frequency dependence differs  strongly from that of a quasi-static experiment. We emphasize that the concept of deviation sensitivity  differs significantly from the tunability \cite{Houssameddine_2008} discussed so far in the context of macrospin oscillators.

\section{Experiment}
Our samples are magnetic tunnel junctions with an ultra-low resistance area product ($0.3~\Omega\mu \mathrm{m}^2$) of the following composition: $IrMn_7 /CoFe_2 /Ru_{0.7} /CoFe_{2.5} /AlOx/CoFe_3/NiFe_5$. The subscripts represent the layers' thickness in $\mathrm{nm}$. They are etched as pillars of $300~\mathrm{nm}$ diameter. In a previous paper \cite{Martin_2011} we have shown that injecting a large dc current through the sample induces the formation of a magnetic vortex in the free layer due to the large Oersted field. At some critical current, spin transfer torque induces  gyrotropic motion of the vortex, thus leading to a large rf response. Its signature is seen in the power spectral density of the junction, that shows a large peak around $400~\mathrm{MHz}$ and up to $10-12$ harmonics. The dynamical properties of this oscillator were thoroughly studied with respect to  synchronization, leading us to deduce it behaves as a  parametric oscillator\cite{Martin_2011}.
Here we examine  frequency modulation on the same samples, in the same experimental conditions. In contrast to other modulation experiments performed on spintronic oscillators \cite{Muduli_2010,Consolo_2010, Pufall_2005}, we sweep the modulation frequency and not the modulation power. The tunnel junction is biased with a large dc current and is simultaneously excited by a small ``low'' frequency current ($\omega_m/2 \pi=3-40~\mathrm{MHz}$) provided by a microwave source. The experiment is realized as follows: the sample is polarized by a dc current through the dc-port of a bias-tee and the ac-port is connected to a power-splitter. The ac-current is delivered by a microwave source to one port of this power-splitter, whereas the power spectral density of the sample is measured on a spectrum analyzer connected to the other port. The amplitude of the dc current used ($I_{dc}$) is typically about $20~\mathrm{mA}$ whereas the amplitude of the ac current ($i_{ac}$) is about $1~\mathrm{mA}$, corresponding  to a frequency deviation of approximatively $15~\mathrm{MHz}$. In the following, the modulation power is corrected using  the experimental attenuation factors.

\begin{figure}[htb]
\includegraphics[width=\columnwidth]{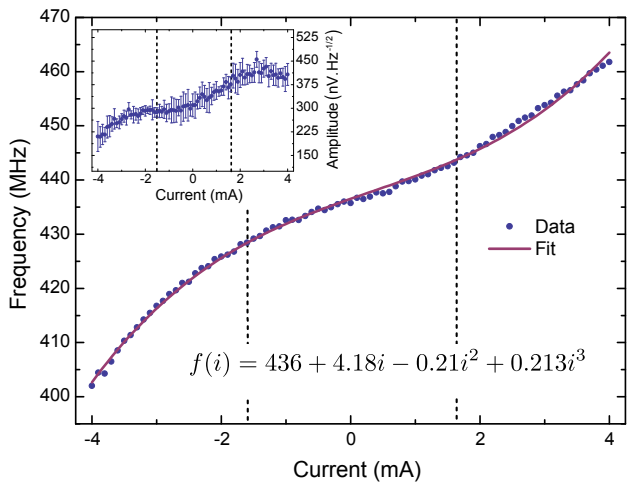} 
\caption{(Color on line) Dependence of the oscillator frequency with bias current around the working point at $I_{dc}=20~\mathrm{mA}$. The solid line corresponds to the polynomial fit. The vertical dashed lines show the maximum current variation imposed by the modulation experiment. Inset: amplitude of the oscillator signal with bias current around the same working point. Fig.~\ref{fig:fI}, \ref{fig:zeros} and \ref{fig:modulation_curves} correspond to the same sample.}
\label{fig:fI}
\end{figure}

\begin{figure}[htb]
\includegraphics[width=\columnwidth]{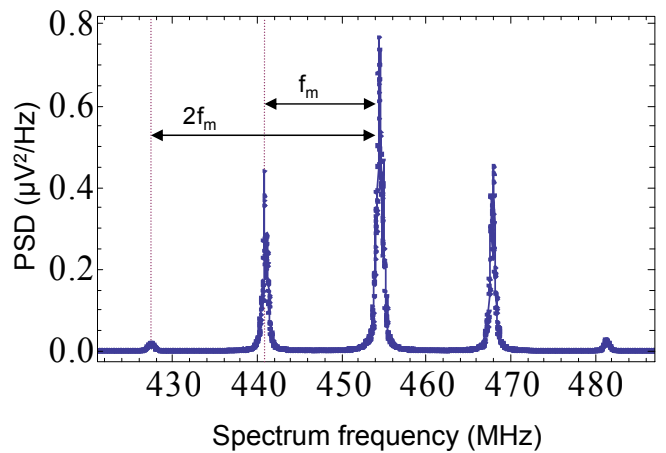} 
\caption{(Color on line) Power spectral density (PSD) of the vortex oscillator subjected to a frequency modulation performed at $f_m=\omega_m/2 \pi=13.5~\mathrm{MHz}$, with a modulating power of $-19.5~\mathrm{dBm}$ corresponding to a current $i_{ac}=0.9~\mathrm{mA} $. The central peak is the carrier and on both sides, first and second order sidebands are clearly observed.}
\label{fig:spectrum}
\end{figure}

\begin{figure}[htb]
\includegraphics[width=\columnwidth]{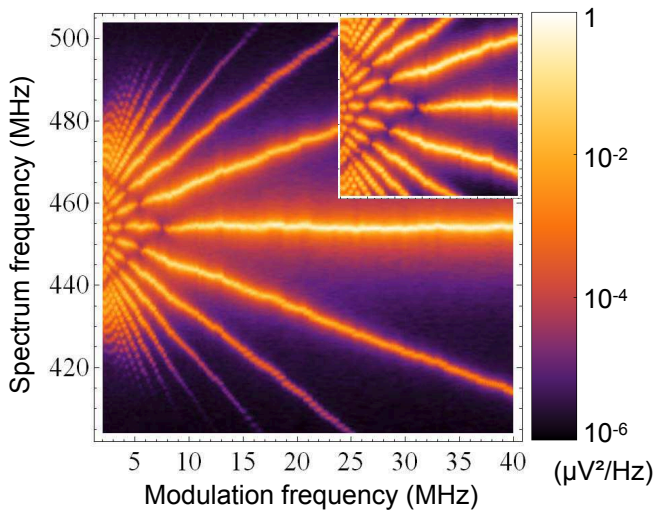} 
\caption{(Color on line) Frequencies of the different peaks, carrier and sidebands, as a function of the modulation frequency. The color scale codes the power spectral density in a logarithmic scale. Inset: zoom on the region of low modulation frequency where the extinctions of the peaks are clearly seen.}
\label{fig:star}
\end{figure}

Before performing the frequency modulation experiment, the oscillator natural frequency dependence with bias current is investigated by measuring the power spectral density for various dc currents (Fig.~\ref{fig:fI}). We observe that the  dependence of oscillator frequency \footnote{In the following, the pulsation $\omega$ is called frequency and its value is given in $\mathrm{MHz}$.} on bias current can be fitted by a third order polynomial:
\begin{equation}
\omega(I_{dc}+i)=\omega_0+a~i +b~i^2+c~i^3
\label{Eq_fI}
\end{equation}
where the higher order terms, with coefficients $b$ and $c$, provide small corrections to  linearity.
This direct relation between the oscillator frequency and the current flowing through the device implies that modulating the microwave current induces a periodic change of the  instantaneous oscillator frequency. On long time scales, such frequency modulation  results in a  spectral power density as in Fig.~\ref{fig:spectrum}.
We observe several peaks: the carrier, at frequency $\omega_c$, close to the natural frequency $\omega_0$ of the oscillator, and several sidebands at frequencies $\omega_l=\omega_c + l~\omega_m$, $l$ being a positive or negative integer. In the following, peaks will  be labeled by the corresponding  $l$: the carrier ($l=0$), the first order sidebands ($l=\pm 1$), the second order sidebands ($l=\pm 2$),... Similarly, sidebands are seen around the carrier harmonics. In this paper, however,  we focus on the frequency range around the fundamental frequency, where the peaks are the largest. The measured spectrum evolves as we sweep up the modulation frequency. Sidebands shift further apart and the peak amplitudes vary: some peaks become larger, others smaller. The total power estimated from the peak amplitudes is observed to be conserved, as expected from frequency modulation theory \cite{Haykin_2009}. This evolution can be observed in the color map (Fig.~\ref{fig:star}) obtained by measuring many spectra at different modulation frequencies: a cross-section of this map is therefore a spectrum like Fig.~\ref{fig:spectrum}. In this color map, peak frequencies are represented as a function of the modulation frequency. The carrier is the horizontal line at $454~\mathrm{MHz}$. The $l$-order sidebands  shift apart from the carrier with a slope $\pm l$. Peak extinctions are easily seen: at a given modulation frequency, a peak disappears completely, eventually the carrier or the $\pm l$ side-bands. Extinctions follow specific pattern, as shown in the inset. We will show that this pattern gives valuable  indications about the parameters that  govern the oscillatory behavior.

\section{Data analysis}
Our experimental results can be analyzed by using the analytical formulation of frequency modulation \cite{VanDerPol_1930, Pufall_2005}.  By taking into account the equation of the frequency dependence with bias current (Eq.~\ref{Eq_fI}), and replacing $i$ with $i_{ac}\cos(\omega_m t)$ in this equation, we obtain the expression of the instantaneous frequency $\omega(t)$. Then the phase $\phi=\int \omega(t) dt$ can easily be  calculated:
\begin{equation*}
\phi=\omega_c t + \frac {B_1}{ \omega_m} \sin(\omega_m t) 
+ \frac {B_2}{ \omega_m} \sin(2\omega_m t)
+ \frac {B_3}{ \omega_m} \sin(3\omega_m t)
\end{equation*}
with 
\begin{eqnarray}
&& \omega_c=\omega_0+1/2~b~i_{ac}^2 \\
&& B_1=a~i_{ac}~+~3/4~c~i^3_{ac} \\
&& B_2=b~i^2_{ac}/4 \\
&& B_3=c~i^3_{ac}/12 
\end{eqnarray}
Knowing that $e^{i z \sin\theta}=\sum_n J_n(z)e^{in\theta}$, the oscillator response in frequency space $V_{max}e^{i\phi}$ can be written as \cite{VanDerPol_1930,Pufall_2005}:
\begin{equation}
\label{Bessel}
V_{max}e^{i \omega_c t}\sum_{n,m,p} J_n(\beta_1)J_m(\beta_2)
J_p(\beta_3)e^{i(n+2m+3p)\omega_m t}
\end{equation}
where $\beta_i=B_i/\omega_m$, $J_k$ are Bessel functions and $V_{max}$ is the signal amplitude of the oscillator without modulation.
Since $\omega_c=\omega_0+1/2~b~i_{ac}^2$, the carrier frequency is usually not  exactly  the natural oscillator frequency, unless  the oscillator frequency depends linearly on bias current. In our case, the deviation of the carrier frequency from the natural frequency is almost  below  experimental accuracy, thus giving another proof of  the quasi-linearity of  frequency with current.
From Eq.~\ref{Bessel}, we see that each peak labeled by the number $l$ has therefore an amplitude equal to a sum of products of Bessel functions, where the indices $n, m, p$ must verify $n+2m+3p=l$.
This infinite sum may be truncated since $J_n(x)$ is negligible when $n>2x$. A very good approximation is, however,  obtained with a much less drastic criterion. With our experimental values of $B_j$, keeping only the terms with $n,m,p \leq 4$  gives a close approximation to the exact solution. 

At this stage, it is worth re-examining  the mathematical formula. In particular, suppose that  $B_2$ and $B_3$ are less than 5\% of $B_1$, which is the case for our vortex oscillator. In that case, Eq.~\ref{Bessel} has the following properties:
i) The extinctions of the carrier, i.e. the zeros of the carrier amplitude function, are  controlled almost exclusively by parameter $B_1$. The impact of the two other parameters on the position of the zeros is below the experimental accuracy. \
ii)  $B_2$ controls the asymmetry between right and left sidebands. If $B_2=0$, the right and left sidebands are equal, whereas when $B_2>0$ (resp. $B_2<0$), the left (resp. right) sideband becomes larger. The effect of $B_2$ on the dissymmetry is linked to the fact that $B_2$ is proportional to the only odd power term in the polynomial expression of $\partial \omega/\partial i$.
iii) $B_3$ mostly modifies the amplitude: for example $B_3>0$ amplifies both first-order sidebands and reduces the second-order. The effect is opposite when $B_3<0$.

\begin{figure}[htb]
\includegraphics[width=\columnwidth]{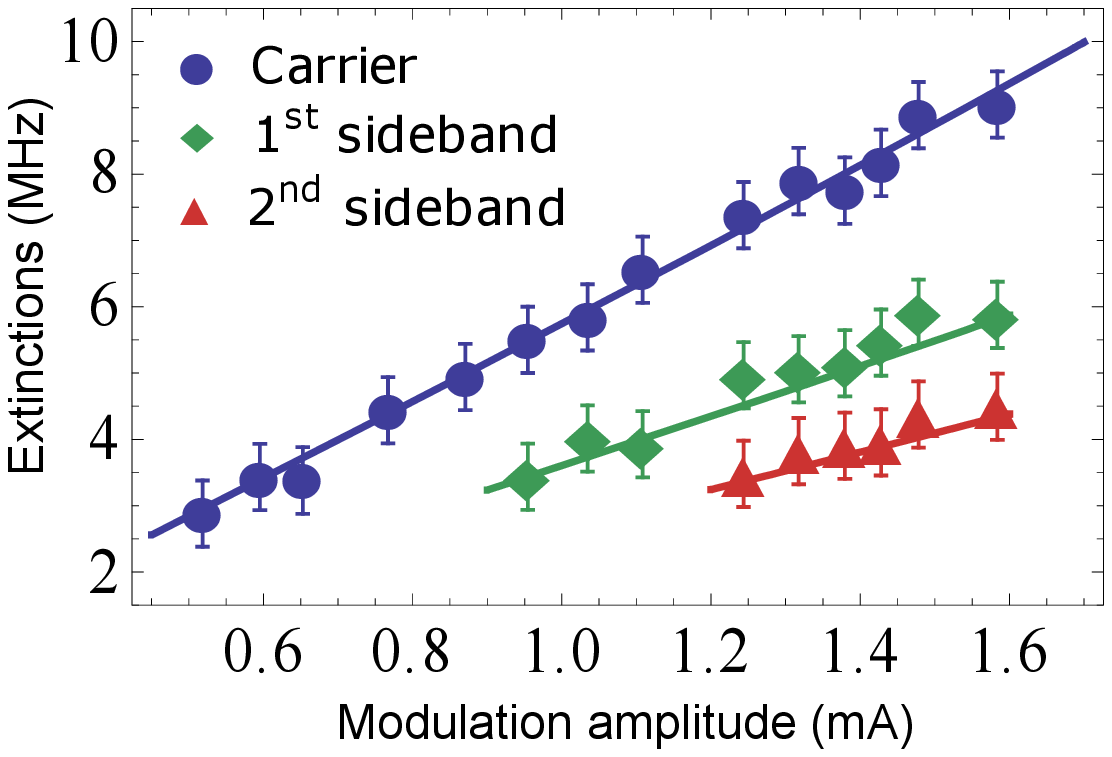} 
\caption{(Color on line) Values of the modulation frequency at which the last extinction is observed, when increasing modulation frequency. Blue dots: extinctions of the carrier; green diamonds: average value of the extinctions of the right and left first order sidebands; red triangles: average value of the extinctions of the second order sidebands. Solid lines correspond to the fit $\omega^*_{m,l}=B_1(i_{ac})/x^*_l$, for $l=0,1,2$.}
\label{fig:zeros}
\end{figure}

\begin{figure}[htb]
\includegraphics[width=\columnwidth]{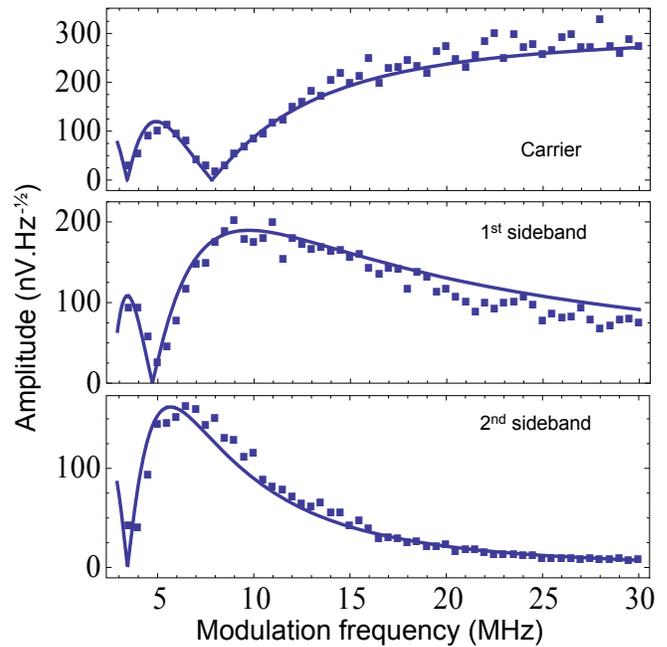} 
\caption{(Color on line) Amplitude of the carrier and of the right sidebands as a function of the modulation frequency, for a modulation current $i_{ac}=1.3~\mathrm{mA}$. The line represents the calculated amplitude using Eq.~\ref{Bessel} with the values of the parameters $a, b, c, V_{max}$ determined by the procedure explained in the text. Similar results are obtained for the left sidebands.}
\label{fig:modulation_curves}
\end{figure}

From these properties, the three parameters $B_1, B_2, B_3$, or equivalently $a, b, c$, can be extracted from the whole set of data, without  fitting  individual curves. Once  $a,~b,~c$ are determined, it is possible to reproduce the amplitude dependence of each peak (carrier, 1st and 2nd order sidebands) with modulation frequency, for each microwave power used, with no adjustable parameter. By contrast, fitting  single curves independently cannot lead to reproducible parameters. In particular, $b,~c$ and $V_{max}$ vary from fit to fit, since they all act  mostly  on the curve amplitude. In this case, $V_{max}$ is even observed to change by 20-30\%, which is not consistent with power conservation. Now let us describe our method in more detail. We know that the carrier extinctions are only controlled by $B_1$. Thus the zeros are necessarily the same as the zeros of $J_0(B_1/ \omega_m)$, since the higher order Bessel functions have a very weak impact on the zeros position. In Fig.~\ref{fig:zeros}, the last carrier extinction $\omega^*_{m,0}$ is plotted as a function of the applied microwave current.
Mathematically $\omega^*_{m,0}$ must verify $B_1/ \omega^*_{m,0}=x^*_0$, where $x^*_0$ is the first zero of the Bessel function $J_0$. Equivalently the extinctions of the $l$ order sidebands correspond to $B_1/ \omega^*_{m,l}=x^*_l$ where $x^*_l$ is the first zero of the Bessel function $J_l$. Since the carrier extinctions are obtained on a larger scale of modulation current, it is  more appropriate  to fit the carrier extinctions to obtain $B_1$ with a reasonable accuracy. Thus the parameters $a$ and $c$ are perfectly determined and  $B_1$ and $B_3$ are fixed for each modulation current.  $b$ is then determined from the sideband asymmetry. To do this, the ratio between the maximum values of the right and left first order sidebands is  calculated numerically as a function of $B_2$. Comparison to the experimental value of the ratio  gives a specific value of $B_2$. By repeating this operation for different applied microwave currents, it is possible to extract $b$ with reasonable accuracy. In our case, we found: $a=13.6~\mathrm{MHz}/\mathrm{mA}$, $b=-1.7~\mathrm{MHz}/\mathrm{mA}^2$, $c=0.23~\mathrm{\mathrm{MHz}/\mathrm{mA}}^3$ and $V_{max}=300~\mathrm{nV}/\sqrt{\mathrm{Hz}}$. With these four parameters, we can reproduce the whole set of data \textsl{i.e.} the carrier ($l=0$) and sidebands ($l=\pm 1,~\pm 2$) amplitudes obtained at $14$ different values of the modulation current, \textsl{ i. e. } $14\times5=70$ curves. The comparison between experimental data and calculated curves is quite satisfactory (see Fig.~\ref{fig:modulation_curves}). Such a success lends confidence  to the extracted parameters.  A more complex approach including amplitude modulation \cite{Muduli_2010, Consolo_2010} is  not necessary here. The reason is  that the oscillation amplitude  depends weakly  on the current in these systems (cf Fig.~\ref{fig:fI}). We also verify \textsl{a posteriori} our initial assumption: for the highest modulation current used, $B_2$ and $B_3$ are smaller than 5\% of $B_1$.  

\section{Discussion}
Finally, let us compare the parameter $a$ extracted from frequency modulation experiment with the value of $a$ obtained from the fit of Fig.~\ref{fig:fI}. Surprisingly, it appears that the two values are quite different. For example, for the sample considered here, the fit of Fig.~\ref{fig:fI} gives $a\approx 4~\mathrm{MHz}/\mathrm{mA}$ whereas the frequency modulation experiment gives $a\approx 13~\mathrm{MHz}/\mathrm{mA}$. The origin of this discrepancy, observed in all samples,  must come  from the dynamics of the experiments. One experiment is performed in a quasi-static regime, whereas the other is performed at a few $\mathrm{MHz}$. In the frequency modulation experiment, the instantaneous frequency is tuned rather quickly in comparison to the  natural oscillator frequency: a period of the modulation cycle corresponds to the time necessary to perform 10 to 100 orbits. It was already shown experimentally \cite{Manfrini_2011} and numerically \cite{Lee_2007} that a vortex cannot immediately jump from one frequency to another. Thus the agility is not infinite and the typical transition time is of the order of $20$ to $80~\mathrm{ns}$. In our frequency modulation experiment, the frequency is continuously varied with a cycle period corresponding to this typical time. The vortex dynamics is therefore expected to be in a transient regime and not in a stationary state. It is then reasonable that the deviation sensitivity $\partial \omega/\partial i_{ac}$ appears significantly different from the tunability  $\partial \omega/\partial I_{dc}$. This difference has never been pointed out so far in spin transfer oscillators \cite{Muduli_2010, Pufall_2005}: frequency modulation data were collected at much lower modulation frequency relatively to the carrier frequency and analyzed using the frequency-current dependence determined with a quasi-static experiment. So, up to now, the concept of deviation sensitivity has been ignored  in the field of spin transfer oscillators. In previous studies, it was reasonable to consider  tunability  as the relevant parameter since the modulation period was much longer than the transition time between stationary dynamical states. \\

In conclusion,  we have shown that four parameters are enough to account for all our frequency modulation experiments. One of those parameters is the deviation sensitivity, which appears to  differ significantly from the tunability measured in a quasi-static regime. Since a modulation experiment consists of  a fast continuous change of states, the characteristic time of the vortex dynamics must be taken into account. In our case, the modulation period approaches the transient time, so the dynamic experiment can no longer be considered as a quasi-static experiment. Hitherto  in spintronic oscillators, the difference between tunability and deviation sensitivity had  not been observed. Here we show the essential difference between these two concepts: a change from a quasi-steady state to  forced transient dynamics. \\


\bibliographystyle{apsrev}

\end{document}